\documentstyle[sprocl,epsf]{article}

\def\be{\begin{equation}}
\def\ee{\end{equation}}
\def\bea{\begin{eqnarray}}
\def\eea{\end{eqnarray}}

\newcommand{\bef}{\begin{figure}}
\newcommand{\eef}{\end{figure}}
\newcommand{\hmp}{ h^{-1}Mpc}
\newcommand{\etal}{{\it et al.}}
\def\spose#1{\hbox to 0pt{#1\hss}}
\def\ltapprox{\mathrel{\spose{\lower 3pt\hbox{$\mathchar"218$}}
 \raise 2.0pt\hbox{$\mathchar"13C$}}}
\def\gtapprox{\mathrel{\spose{\lower 3pt\hbox{$\mathchar"218$}}
 \raise 2.0pt\hbox{$\mathchar"13E$}}}
\def\inapprox{\mathrel{\spose{\lower 3pt\hbox{$\mathchar"218$}}
 \raise 2.0pt\hbox{$\mathchar"232$}}}

\begin{document}

\title{Scale invariance in galaxy structures:
 general concepts and application to the most recent data
 \footnote{ In the proceedings of the 
"VI Colloque de Cosmologie" 
Paris 4-6 June 1998}}

\author{Luciano Pietronero$^1$ and Francesco Sylos Labini$^{1,2}$}

\address{      	
		$^1$INFM Sezione Roma1,        
		Dip. di Fisica, Universit\'a "La Sapienza", 
		P.le A. Moro, 2,  
        	I-00185 Roma, Italy. 
        	\\
        	$^2$D\'ept.~de Physique Th\'eorique, Universit\'e de Gen\`eve,  
		24, Quai E. Ansermet, CH-1211 Gen\`eve, Switzerland.
		}

\maketitle
\abstracts{The debate on the correlation properties
of galaxy structures has having an increasing interest
during the last year. In this lecture we 
discuss the claims of different authors
who have criticized our approach and results.
In order to have a clear cut of the 
situation, we focus mainly on galaxy distribution 
in the intermediate range of distances $\sim 100 \div 200 \hmp$.
In particular we discuss: (i) the validity of the actual data
and the concept of {\it fair sample}, (ii) the shift of $r_0$
with sample depth and luminosity bias, (iii) 
the value of the fractal dimension, (iv)
the problem of the counts
from a single point and the case of ESP, (v) uniformity of angular catalogs.
The detection of fractal behavior  up to $\sim 100 \div 200 \hmp$
is enough to rise serious problems
to the usual statistical methods used for the characterization
of galaxy correlations,   
standard interpretation of galaxy distribution and
 theoretical models developed.
The clarification of the intermediate  scale behavior
is very instructive for the subsequent interpretation 
of the very large scale galaxy distribution.
}

\section{Introduction}

The presence of large scale structures 
(hereafter LSS)  of galaxies
and galaxy clusters is nowadays one of the most
intriguing feature of the visible universe. The discovery
of the inhomogeneous galaxy distribution
began with the availability of 
intense redshift measurements in the eighties. 
The statistical characterization of such 
structures is however a matter of debate. 
While there is a general agreement that
the inhomogeneities seen in the galaxy catalogs
correspond to a correlated fractal distribution
up to $10 \div 20 \hmp$, at larger scales
the correct statistical properties
are still controversial. 
Moreover there is not  agreement on the 
value of the fractal dimension even at small scales.
The clarification of these facts is crucially
important with respect  to the theoretical interpretation
of such structures. In fact, homogeneity of  matter
distribution is one of the cornerstone of the
Hot Big Bang model as well as of
the different
models of galaxy formation,
and then it is quite important
to establish whether galaxy structures
approach  to a smooth distribution on large enough
scales.

By the analysis  of
galaxy redshift samples one can study
the structure of the Local Universe
up to $\sim 100 \div 200
 \hmp$. The recent availability
of wide angle  galaxy surveys
such as CfA, SSRS, Perseus-Pisces, 
APM-Stromlo, IRAS and LEDA, has in fact allowed a 
statistical description 
of the nearby luminous matter distribution
in a rather complete way. The fact that galaxy redshift 
catalogs
are dominated by LSS and huge voids 
has lead   several authors 
to define them {\it not fair}
with respect to the {\it expected}  statistical
properties of the large scale
matter distribution in the universe, i.e.
homogeneity  \cite{pee93,dac94}.
By using the standard two-points correlation function
it has been found \cite{dp83,pee93}
that galaxy structures are characterized 
by having a very well defined "correlation length"
that is established to be $r_0 \approx 5 \hmp$.
The physical interpretation of such a scale
being the distance above which
the density fluctuations become
of the same order of the average density. At twice this
distance the fluctuations have
small amplitude and the linear theory holds,
i.e the distribution becomes homogeneous.
The evidence in favor of this result coming
from various different observations:
the uniformity of galaxy angular
catalogs, the number counts of galaxies as a function
of apparent magnitude and  
the so-called "rescaling" of the amplitude of the 
angular correlation function.
In addition, it has been found by several
authors \cite{dav88,park94,ben96} 
that the correlation
length $r_0$ has a  very qualitative dependence on
galaxy luminosity. According to such an effect,
called "luminosity bias",  
$r_0$ increases in the samples
which contain the brightest 
galaxies.  
In summary, the most popular point of view, is   that
galaxy distribution
exhibits indeed fractal properties
with dimension $D \approx 1.3$ at  small
scale (i.e. up to $\sim 10 \div 20  \hmp$), and at 
larger scales there is an overwhelming
evidence in favor of homogeneity.
We will refer to this 
{\it picture of galaxy clustering as the 
standard one. }
The main problems that this picture rises are:
\begin{itemize}
\item The problem of the statistical validity of different
redshift samples.
\item The presence of LSS and the detection of a small correlation length.
\item The mismatch galaxy-cluster (different correlation lengths
of galaxies and galaxy clusters).
\item The luminosity bias and the clustering properties of different
galaxy types.
\item The possible {\it "reconstruction"} of the three
dimensional properties from angular catalogs.
\end{itemize}

This problematic situation has lead us to reconsider the basic
assumption of the usual analysis of galaxy clustering.
In this respect, a completely different interpretation
of galaxy correlations has been suggested
by our group
\cite{pie87,cp92,slmp98}.
By analyzing the different galaxy samples
with the statistical tools
suitable and appropriate to characterize
highly irregular distributions, 
as well as regular ones, it has been
found that
\begin{itemize}
\item  galaxy
and cluster  distributions exhibit
very well defined scale invariant
properties with fractal dimension
$D \approx 2 $ up to the limits
of the analyzed sample ($R_s \approx 150 \hmp$),
\item   the different catalogs show
 almost the same statistical properties
and are in a reasonable agreement
with each other,
\item  the standard picture of galaxy correlations
has been derived by an analysis which assumes 
a priori  homogeneity of matter distribution.
Such a statistical method is not suitable
for the characterization of self-similar structures.
In particular, both the fractal dimension
($D \approx 1.3$) and the "correlation length"
$r_0 \approx  5 \hmp$ are artifact of
the data analysis, and do not correspond
to the correct statistical properties 
of galaxy distribution. The standard correlation
analysis is one of the statistics usually used,
which are not suitable to detect the 
properties of irregular (self-similar) 
structures. Others are the power spectrum,
the density contrast  as
a function of scale  and quantities related
(see Sylos Labini \etal 1998\cite{slmp98}
 - hereafter Paper 1). 
\item There are various evidences, which
are statistically weaker, and 
which support the continuation of the fractal
behavior with $D \approx 2$ up to the deepest
scale investigated so far. 
\end{itemize}


Mandelbrot \cite{man77} was the first to propose
the relevance of fractal structures in respect 
with the problem of galaxy clustering.
He has introduced the concept of "Fractal Geometry" 
which has opened a new perspective on natural phenomena
and of which we will make extensive use in what follows.
However up to the eighties the discussion
was based mainly on the angular data, as the redshift measurements
were very sparse.
The debate on galaxy correlations has had an important 
impulse in 1987 with the work of L.P. \cite{pie87}.
Then in 1996 in the Conference "Critical Dialogues in Cosmology"
there has been
a debate between Prof. Marco Davis and one of us (L.P.)
\cite{dav97,pmsl97}.  
The point of view of Prof. Davis was that $r_0$ 
is indeed a real correlation length and hence a
characteristic scale of galaxy distribution, while
L.P. has argued in favour of a fractal behavior
of galaxy distribution {\it in the available samples}.
 In the last few months
a wide debate on this subject is
in progress 
\cite{gu97,slmp98b,coles98,scara98,cappi98,joyce98,man98,wu98,pee98} 
and different authors
have expressed different points of view,  which are  in general
more articulated with respect to the one of Davis.
(See the web page {\it
 http://www.phys.uniroma1.it/DOCS/PIL/pil.html}
 where all these materials have been collected).
However different authors
supporting the homogeneity of large scale 
galaxy distribution are often in contradiction with each other.
In this lecture, after reviewing our main results,
we address the most controversial
points in this debate, which can be summarized as follows: 

\begin{itemize}

\item Validity of the actual data 
and the concept of {\it "fair sample}.

\item The shift of $r_0$ with sample depth and
luminosity bias.

\item Value of the fractal dimension.

\item Problems of counts from a single point:
radial counts in the European Slide Project (ESP)
 redshift survey.

\item Uniformity of the angular catalogs.

\end{itemize}
 In our opinion the fact that the clarification of the 
 statistical properties of the local universe (i.e. up to
 $\sim 100 \hmp$) is a fundamental fact with respect 
 to the interpretation of the far away galaxy distribution.
 (We note that in this range of scale the data are statistically
 very robust.)
 Instead of reviewing the different arguments 
 for scales larger than $\sim 100 \hmp$ (see Paper 1\cite{slmp98})
 we will focus 
 our discussion on smaller scales.
 Moreover 
 we briefly discuss the fact that the CDM-like models
 of galaxy formation are unable to reproduce the
 clustering properties of real galaxy samples.
 The problem here is twofold. From one side the 
 theoretical models have been developed to explain
 the incorrect properties deduced from an inconsistent analysis
 of galaxy surveys. From the other side
 the simulations 
 are analyzed, as well as real galaxy surveys,
 with   statistical methods which are unable to 
 recover the properties of irregular (self-similar)
 distributions. 
 From this confusing situation several concepts have arisen,
 like the bias galaxy formation, the luminosity bias 
 and others, which are not appropriate to describe the
 scale-invariant structures observed in the data.
 
 A interesting theoretical attempt in the new perspective
 of scaling exponents have been developed
 very recently the group of Prof. N.Sanchez and Prof. H.
 De Vega\cite{sanchez1,sanchez2}.
 We stress below the importance of
 the change of theoretical perspective
 related to the use of the proper 
 statistical methods.


 \section{Review of main results and statistical validity of galaxy data}

 The proper methods to characterize irregular as 
 well as regular distributions have been
 discussed    in Coleman \& Pietronero\cite{cp92}
 and Sylos Labini \etal \cite{slmp98} 
 in a detailed and exhaustive way.
 The basic point is that, as far as a system 
 shows power law correlations, the usual $\xi(r)$ analysis 
\cite{pee80} gives an incorrect result, since it is 
 based on the a-priori assumption of homogeneity. 
 In order to check whether homogeneity is present 
 in a given sample one has to use 
 the conditional density $\Gamma(r)$ 
 defined as \cite{pie87}
 \be
 \label{eq1}
 \Gamma(r) = \frac{\langle n(r_*) n(r_*+r) \rangle}{\langle n 
 \rangle} =
 \frac{BD}{4 \pi} r^{D-3}
 \ee
 where the last equality holds in the case of a fractal
 distribution
 with dimension $D$ and prefactor $B$. 
 In the case of an 
 homogenous distribution ($D=3$) the conditional density 
 equals the average density in the sample. 
 Hence the conditional density is the suitable 
 statistical tool to identify 
 fractal properties (i.e. power law correlations
 with codimension $\gamma=3-D$) as well as 
 homogeneous ones (constant density with sample size).
 If there exists a transition scale  $\lambda_0$ 
 towards homogenization,
 we should find $\Gamma(r)$ constant for scales 
 $r \gtapprox \lambda_0$.

 It is simple to show that 
 in the case of a fractal distribution the 
 usual $\xi(r)$ function in a spherical sample of radius $R_s$ 
 is \cite{pie87,cp92}
 \be
 \label{eq2}
 \xi(r) = \frac{D}{3} \left( 
 \frac{r}{R_s} \right)^{D-3} -1 \; .
 \ee
 From Eq.\ref{eq2} we can see two main problems of the 
 $\xi(r)$ function: its amplitude depends on the sample size $R_s$
 (and the so-called correlation length $r_0$, defined as 
 $\xi(r_0) \equiv 1$, linearly depends on $R_s$) 
 and it {\it has not} a power law behavior.
 Rather the power law behavior is present 
 only at scales $r  \ll r_0$, and 
 then it is followed by a sharp break in the log-log plot
 as soon as $\xi(r) \ltapprox 1$. Such a behavior does not
 correspond  to any real change of the correlation 
 properties of the 
 system (that is scale-invariant by definition) and it 
 makes extremely difficult the estimation
 of the correct fractal dimension as it is 
 shown in Fig.\ref{fig1}. 
 In particular if the sample size is not large enough 
 with respect to the 
 actual value of $r_0$, 
 the codimension estimated by the $\xi(r)$ function 
 ($\gamma \approx 1.7$) 
 is systematically larger than $3-D$ ($\gamma \approx 1$) 
 \cite{slmp98}.
 \bef
 \epsfxsize 6cm 
 \centerline{\epsfbox{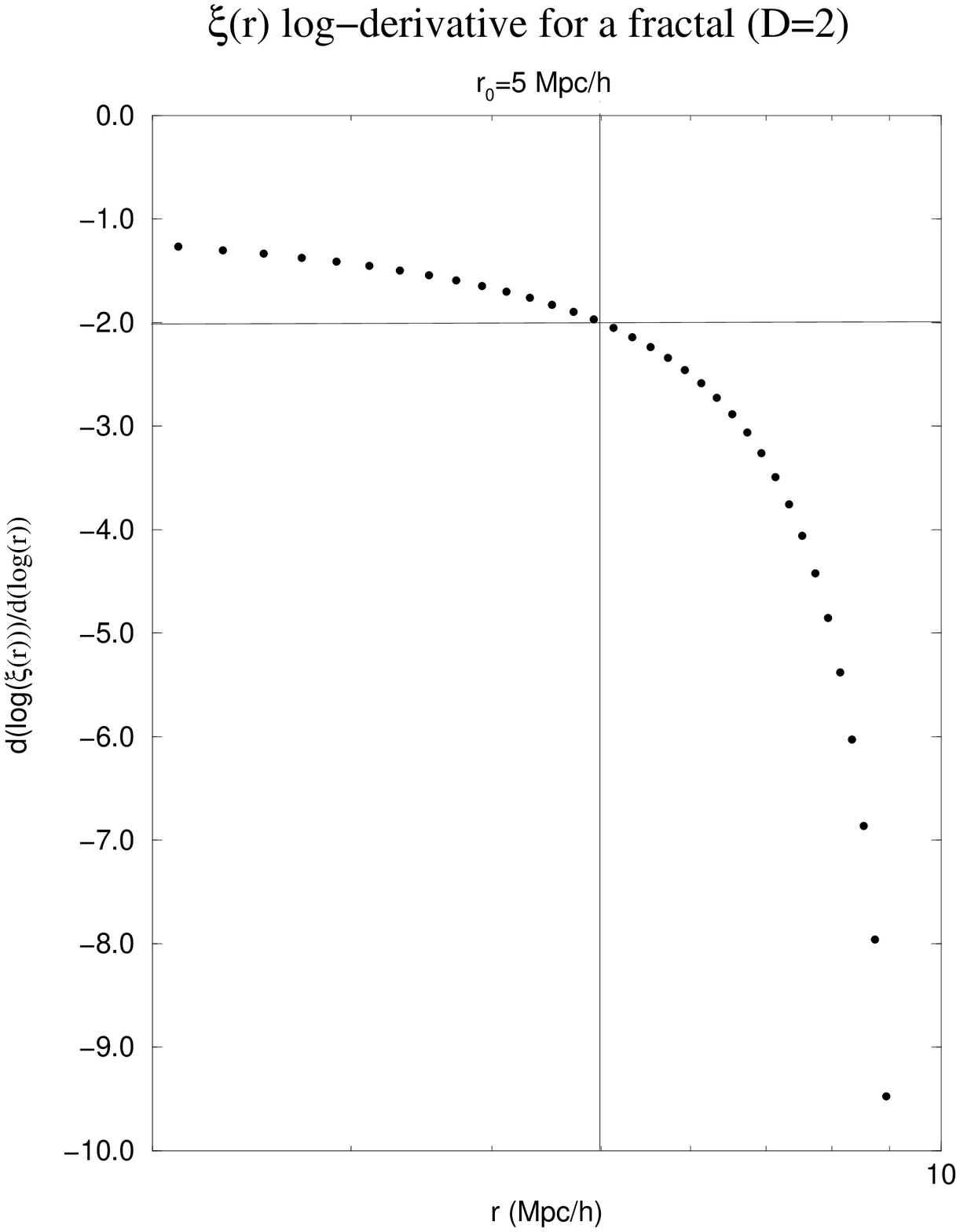}} 
 \caption{\label{fig1} 
 Behaviour of the tangent to $\xi(r)$
 for a mathematical fractal with dimension $D=2$
 in the log-log plot. The sample depth $R_s$ is such that
 $r_0=5 \hmp$. We may see that at $r=r_0$ the slope is $-2\gamma$
 and then it rapidly decays to $-\infty$.}
\eef

 Given this situation it is clear that the $\xi(r)$ analysis 
 is not suitable
 to be applied unless
 a clear cut-off towards homogenization 
 is present in the samples analyzed. As this is not the case, 
 it is appropriate and convenient to use 
 $\Gamma(r)$ instead of $\xi(r)$. We have discussed 
 in detail in Paper 1 that the use of the 
 correct statistical methods is complementary 
 to a change of perspective from a theoretical
 point of view. 
 In Tab.\ref{tab1} we report the characteristics of the 
 various catalogs we have analyzed by using
 the methods previously illustrated.
 \begin{table}
 \caption{\label{tab1}The volume 
 limited catalogues are characterized by the following 
 parameters:
 - $R_d (\hmp)$ is the depth of the catalogue
 - $\Omega$ is the solid angle
 - $R_s (\hmp)$ is the radius of the largest sphere 
 that can be contained in the catalogue volume. 
 This gives the limit of statistical validity of the sample.
 - $r_0(\hmp)$  is the length at which $\xi(r) \equiv 1$.
 - $\lambda_0$ is the eventual real crossover to a homogeneous 
 distribution that is actually never observed. 
 The value of $r_0$ 
 is the one obtained 
 in the deepest VL  sample. 
 (Distances are expressed in $\hmp$).
 }
 \begin{tabular}{|c|c|c|c|c|c|c|}
 \hline
     &      &          &    &              &  &       \\
 \rm{Sample} & $\Omega$ ($sr$) & $R_d  $ & $R_s  $ 
 & $r_0  $& $D$ & $\lambda_0 $ \\
     &       &    &    &    &              &       \\
\hline
CfA1         & 1.83      & 80 & 20 & 6  & $1.7 \pm 0.2$ & $ >80 $    \\
CfA2South    & 1.23      & 130& 30 & 10 & $2.0 \pm 0.1$ & $ >120$    \\
PP           & 0.9       & 130& 30 & 10 & $2.0 \pm 0.1$ & $ >130$    \\
SSRS1        & 1.75      & 120& 35 & 12 & $2.0 \pm 0.1$ & $ >120$    \\
SSRS2        & 1.13      & 150& 50 & 15 & $2.0 \pm 0.1$ & $ >130$    \\
Stromlo-APM  & 1.3       & 100& 35 & 12 & $2.2 \pm 0.1$ & $ >150$    \\
LEDA         & $\sim 5 $ & 300& 150& 45 & $2.1 \pm 0.2$ & $ >150$    \\
LCRS         & 0.12      & 500& 18 & 6  & $1.8 \pm 0.2$ & $ >500$    \\
IRAS$2 Jy$   & $\sim 5$  & 60 & 20 & 5  & $2.0 \pm 0.1$ & $ >50$     \\
IRAS$1.2 Jy$ & $\sim 5$  & 80 & 30 & 8  & $2.0 \pm 0.1$ & $ >50$     \\
ESP          & 0.006     & 700& 8  & 3  & $2.0 \pm 0.2$ & $ >700$    \\
             &           &         &    &               &    &       \\
\hline
\end{tabular}
\end{table}
Let us consider in more detail
some recent catalogs and discuss the  disagreement of Tab.1 with the 
analyses reported by several other authors \cite{gu97,lov96,lcrs}

\begin{itemize}

\item ESP. In this case  the estimation of $R_s$ slightly differs
from that of Guzzo\cite{gu97},
 probably because we\cite{slmp98} have not used 
the relativistic corrections 
(see below). Also
the value of $r_0$ is slightly different 
($r_0 \sim 3 \hmp$ instead of
the measured $r_0 \sim 4.5 \hmp$), and this is probably due 
to the fact that ESP does not cover a continue solid angle in the 
sky, as it is a collection of pencil beams. Such a 
situation necessarily requires the introduction of 
spurious treatments of the boundary conditions:
we limit the analysis of the conditional density 
to a depth $R_s$ corresponding to the radius
of the maximum sphere fully contained in the sample volume.

\item LCRS. This survey has the peculiar property of being 
limited by two limits in apparent magnitude (a lower and 
an upper one). In order to construct a VL sample
in this case,
 one has to impose two limits in distances and
correspondingly two in absolute magnitude. This
is the origin of a smaller $R_s$ 
in our Table 1 than this reported by Guzzo ($R_s \sim 32 \hmp$).
 This implies a smaller $r_0$,
 much closer to the measured one.

\item Stromlo/APM. We have extensively analyzed this catalog 
 \cite{slmp98,slm98} and the value of
$r_0$ is reported in Tab.1. 
Due to the sparse sampling strategy 
adopted to construct this catalog,
we are able to measure the correlation properties up to
$R_s \sim 40 \hmp$ and not $83 \hmp$ as reported by Guzzo\cite{gu97}.
The disagreement with the work 
of Loveday \etal \cite{lov96} ($r_0 \approx 12 \hmp$ rather than
$r_0 \approx 5 \hmp$) is probably due to the treatment of
the boundary conditions and their use of ML samples rather than 
VL ones (i.e. they used weighting schemes with 
the luminosity selection function). In any case, we stress again,
the proper test is to 
check whether the conditional density has 
a power behavior \cite{slm98}.
\end{itemize}

 We show in Fig.\ref{fig2} the results
 of the conditional density
 determinations in various redshift surveys \cite{slmp98,mslgpa97}.
 All the available data 
 are consistent with each other and show fractal correlations with
 dimension $D = 2.0 \pm 0.2$ up to the deepest scale probed up to
 now by the available redshift surveys, i.e. $\sim 150 \hmp$.
 A similar result has been obtained by the analysis of galaxy cluster
 catalogs \cite{msla97,slmp98}.
 \bef 
 \epsfxsize 10cm 
 \centerline{\epsfbox{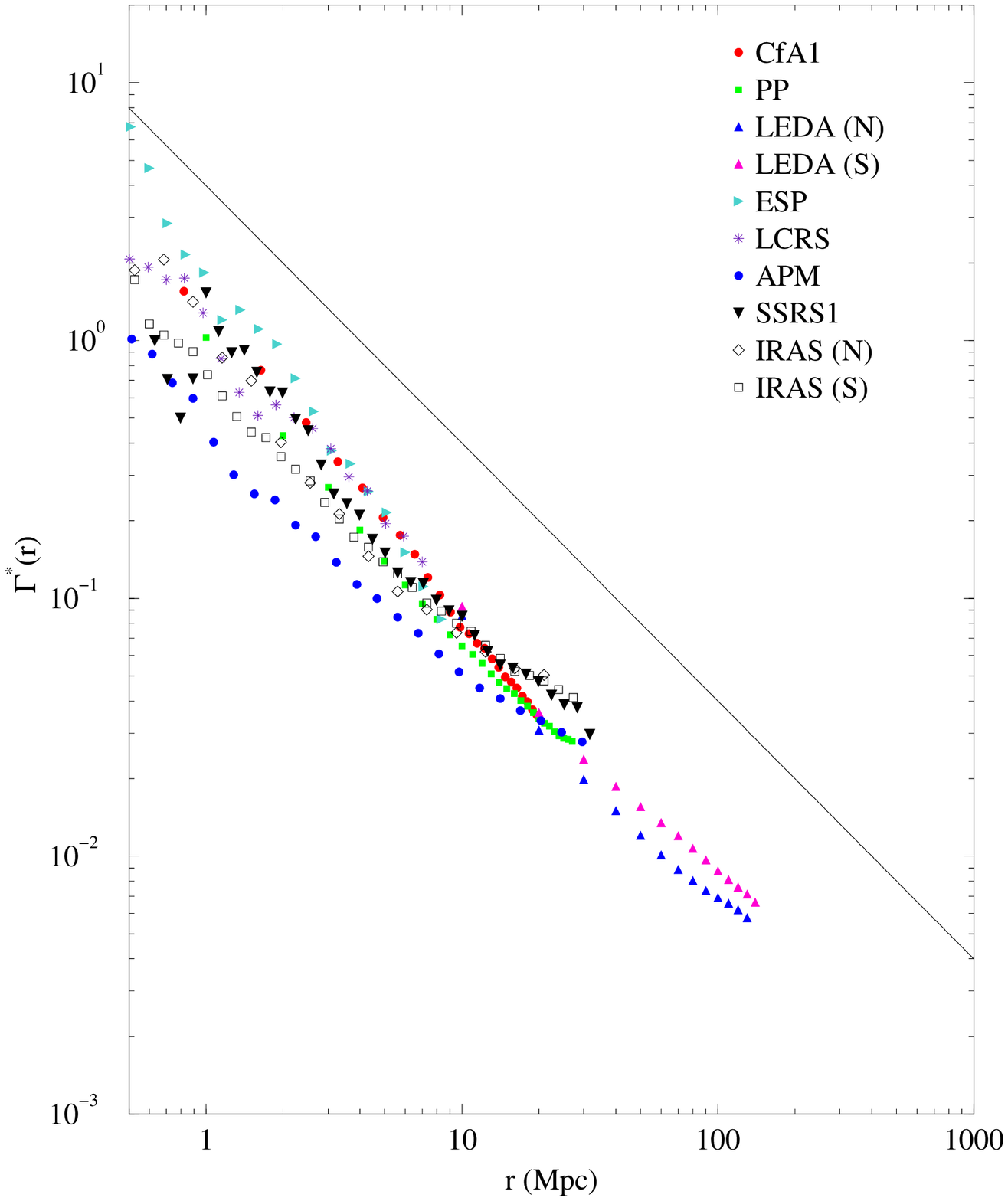}}  
 \caption{\label{fig2} 
 The spatial density   $\Gamma(r)$     computed in some 
 VL samples of CfA1, PP, LEDA, APM, ESP, LCRS, SSRS1, IRAS 
 and ESP (from Sylos Labini \etal, 1998). } 
 \eef


\section{Shift of $r_0$ and luminosity bias}

 Several authors \cite{gu97,wu98,cappi98}  
 based a part of 
 their arguments if favour of homogeneity 
 on the fact that a luminosity bias is responsible
 of the shift of $r_0$ with sample size\cite{dav88,dav97}.
 In this respect, we would like to remark that  
 the authors  \cite{park94,ben96} 
 who have addressed  this 
 concept  have  never presented {\it any quantitative argument
 that explains the shift of $r_0$ with sample size.}
 An exception is represented by the 
 paper of Davis \etal \cite{dav88}
 in which the authors claim that
 $r_0$ behaves as $R_s^{0.5}$. However for what concerns
 luminosity segregation, the meaningful parameter must be the absolute 
 magnitude limit of the volume limited (VL)
 sample considered rather than the depth ($R_s$).
 Having fixed the limiting apparent magnitude of the 
 catalog, at each $R_s$ would correspond a well defined 
 absolute magnitude. The brightest galaxies are  present yet
 in samples like CfA1, and hence according to
 the luminosity segregation paradigm, there is no reason  one should 
 expect that in deeper sample (like CfA2 or SSRS2) $r_0$
 is increased. However this is actually the case 
 \cite{ben96,park94}.

 We have discussed in detail in \cite{slp96} 
 that the observation
 that the giant galaxies are more clustered
 than the dwarf ones, i.e. that the massive elliptical
 galaxies lie 
 in the peaks of the density field, is a consequence of the
 self-similar behavior of the whole {\it matter distribution}. The increasing
 of the correlation length of the $\xi(r)$ has
 nothing to do with this effect, rather it is
 related to the sample size.
 It may be useful to mention that the view presented here actually
 redefines the notion of bias.  In standard biasing scenarios the 
 distribution of  galaxies of different luminosity 
 are  Gaussian but 
 with different amplitudes. Here we consider the fact that  
 galaxies of different luminosities  
 have different correlation
 properties and hence different fractal dimensions (the cluster
 distribution being a coarse graining of the galaxy one, and hence having the same
 fractal dimension $D\approx 2$ of that of galaxies).
 The more detailed picture  is just that the fractal
 dimension of a set of density fluctuations  depends on the 
 threshold. This tendency is actually indicated by present observations,
 which show a slight increase of the fractal dimension with decreasing
 absolute luminosity of the galaxies in the sample \cite{slp96} 
 (however,
 still with relatively modest statistics). 
 Such a scenario is naturally
 formulated within the framework of multi-fractals \cite{pie87,slmp98}.

It remains open the question whether the distribution of dark matter
continues the multi-fractal behaviour found in the galaxies.
In other words, it would be possible that the dark matter 
 fills homogeneously the space:
in this case the fractal dimension of the distribution of
dark matter would be $D=3$. Such a scenario has been 
studied by Durrer \& Sylos Labini \cite{durr98} 
and Baryshev \etal\cite{bar98} in more detail.
 
 
\section{Value of the fractal dimension }

 On the actual value of the fractal dimension 
 at small scale ($< 10 \hmp$) there are, at least, 
 three different
 points of view. The first one is the canonical 
 point of view of Peebles \cite{pee93,pee98},
 who obtains from the 
 standard correlation function $\xi(r)$ a value
 of $D \simeq 1.3$ in the range of scale $0.1 \hmp \ltapprox r
 \ltapprox 10 \hmp$. The second is due to Guzzo \cite{gu97} who 
  claims that $\Gamma(r)$  is a power law with $D \sim 1.2$ up to
 $ r \sim 3.5 \hmp$, then it shows  a different scaling  
 between $\sim 3\hmp$ and $30 \hmp$ with fractal 
 dimension $D\approx 2$.  
 Finally we claim that $D\approx 2$ in the whole 
 range of scale $0.5\hmp \ltapprox r \ltapprox 150 \hmp$
 (This result has been recently confirmed by Cappi \etal\cite{cappi98}
 by the analysis of the SSRS2 galaxy sample.)
 The fact that the first two determinations are in contradiction 
 has never been discussed by the authors \cite{gu97,pee98,wu98}.
 The first point has been discussed in Sec.2 and here we have nothing new
 to add. We focus on the determination of the fractal dimension at very 
 small scale $ r < 10 \hmp$ by the conditional density analysis,
 clarifying the difference with our result with that of Guzzo and
 coworkers \cite{guzzo91}.

Suppose, for simplicity,  we have a 
spherical sample of volume $V$ in which there 
are $N$ points, and we want to measure the conditional density.
It is possible to compute  the average  distance 
between neighbor galaxies $\langle \Lambda \rangle$, in a fractal
distribution with dimension $D$, and the result is 
\be
\label{e5}
\langle \Lambda  \rangle = \left(\frac{1}{B}\right)^{\frac{1}{D}} \Gamma
\left(1 + \frac{1}{D} \right)
\ee
where $\Gamma$ is the Euler's gamma-function \cite{slm98}.
(Note that the prefactor $B$ is dependent on the luminosity
selection function of the VL chosen).
Clearly this quantity   is related 
to the lower cut-off of the distribution $B$ (eq.\ref{eq1}) 
and to the fractal dimension $D$.
If we measure the conditional density at distances 
$ r \ltapprox \langle \Lambda  \rangle$, 
we are affected 
by a {\it finite size effect}. 
In fact, due the depletion of points at these 
distances 
we underestimate the real conditional density finding an higher value 
for the correlation exponent (and hence a lower value for the fractal 
dimension). 
In the limiting case, for distances $ r \ll \langle \Lambda  \rangle $, 
we   find almost no points and 
the slope is
$\gamma=-3$ ($D=0$). 
In general, when one  measures $\Gamma(r)$ at distances 
which correspond to 
a fraction of $ \langle \Lambda  \rangle$, 
one finds systematically an higher value of the 
conditional density exponent. 
Such a trend  is completely spurious and due to the depletion of
points at such distances. It is worth to notice that this effect
gives rise to a curved behavior of 
$\Gamma^*(r)$  
(the integral of $\Gamma(r)$\cite{cp92,slmp98})
at small distances, because 
of its integral nature. This is exactly the case of the deepest VL 
of Perseus-Pisces which Guzzo \etal  \cite{guzzo91} considered in their analysis,
and for which $ \langle \Lambda  \rangle  \sim 8 \hmp$. 
Note that $\langle \Lambda  \rangle$ is of order $1 \hmp$
for a VL sample with $M_{lim} \approx -18$ while it growths
up to $\sim 10 \div 15 \hmp$ for VL samples with $M -20 \div -21$.
This is due to  the depletion of points 
in VL samples with brighter magnitudes\cite{slmp98}.

A clarifying test 
in this respect would be to check whether this change of slope
is actually present also in the others VL samples of the same survey, 
which have a larger number of points (and hence a lower $\langle \Lambda 
 \rangle$).
This test has been performed  by our 
group\cite{slmp98} and the conclusion is
that the change of slope is {\it due a finite size effect} rather
being an intrinsic property of galaxy distribution.

\section{Problems of counts from a single point and the case of ESP}

In a recent paper Scaramella \etal\cite{scara98} 
(hereafter S\&C ) have applied
the same statistical analysis to a deep 
survey  of large scale structure -the ESP galaxy survey -
as performed by us 
(Paper 1).
Despite the adoption of the same method, they 
have reached very different conclusions: That there is no
evidence for a scale invariant distribution of galaxies
with dimensionality $D \approx 2$, as argued in Paper 1,
and that their results favour the ``canonical''
value of $D=3$ (corresponding to a homogeneous universe).
In Joyce \etal\cite{joyce98} we have discussed
in more detail this matter: here we review the main results.

As already mentioned 
the length scale for analysis with the conditional density,
is fixed 
by the maximum radius of a sphere which can be fitted inside
the sample volume of the corresponding surveys, since the 
statistic $\Gamma(r)$ can only be computed up to such a distance.
To extract information from surveys 
about correlations
at length scales greater than this, one needs to consider
other statistics. The most simple one is the radial count 
$N(<r)$ from the origin in whatever solid angle is covered 
by the survey. Depending on the survey geometry the difference
between the length scales to which we can calculate $\Gamma(r)$
and $N(<r)$ varies greatly. The number count from the origin
is obviously a much less powerful statistic since it doesn't 
involve the average. It is intrinsically a more fluctuating 
quantity. Such fluctuations are, however, about the average 
behaviour and,  by sampling a sufficiently large range of scale,
one should be able to recover the average behaviour of the 
number count $\sim r^D$. What one means by ``sufficiently large range
of scale'' depends
strongly on the underlying nature of the fluctuations.
In particular (see section 6.1 of Paper 1) the
cases of a homogeneous distribution with Poissonian type
fluctuations and a fractal structure with scale invariant 
fluctuations are quite different. In the former case one 
predicts a very rapid approach to perfect $D=3$ behaviour
at a few times the scale characterizing the fluctuations;
in the latter $N(<r)$  has intrinsic fluctuations
on all scales (which average out in $\Gamma(r)$), 
in addition to the statistical 
sampling Poisson noise in $N(<r)$ which dominates
up to a scale which depends on the sample (on the number of points,
and therefore on the solid angle of the survey). 

We now turn to the ESP survey, which is a very deep survey 
extending to red-shifts $z \sim 0.3$ in a very narrow
solid angle $\Omega \approx 0.006 \; sr$. Because of this geometry 
the analysis with $\Gamma(r)$ only extends to $R_s \approx 10 \div 12$;
in this regime it shows a clear $D\approx 2$ behaviour
consistent with the other surveys analyzed in Paper 1.
With the radial counts from the origin $N(<r)$, however,
the analysis can extend to distances almost two orders of 
magnitude greater. The results of the latter analysis
can be summarized as follows:

$\bullet$ Up to a scale of $\sim$ 300 $\hmp$ the number counts are
highly fluctuating;

$\bullet$ Beyond this scale the counts can be well fitted by a 
fairly stable average behaviour  $N(<r) \sim r^D$. 
The value 
obtained for the dimension of the number count $D$ 
in this range depends (i) on the assumptions about the
cosmology and (ii) on the so-called K corrections.
The uncorrected data (i.e. with the euclidean distance 
relation and without K corrections) gives a slope of 
$D \approx 2$, while both corrections lead to an increase in the slope.
In particular S\&C  apply the corrections in a way
which produces  values $D \approx 3$ which they argue to
reflect the true behaviour of the galaxy distribution.
  
Consider first what conclusions may be drawn from the
fluctuating regime. If the universe is homogeneous on 
large scales, the data clearly show that the scale 
characterizing such homogeneity is at least
$\sim 100 \div 300 \hmp$.
The implication from our discussion above is that
any standard analysis using the correlation function
is inappropriate for characterizing the properties of
the fluctuations.  In particular the authors claim 
in various papers   that
in the ESP galaxy sample, $r_0 \approx 4 \hmp$.
The origin of this length scale can also be
inferred from the discussion above:
In a fractal distribution (with $D=2$) we see from
(\ref{eq2}) that $r_0 \approx R_s/3$, and as 
noted above $R_s \approx 10 \div 12$ in the ESP
survey. It is related not to a physical
property of the galaxy distribution but to the
specific geometry of the survey it has been
determined from. The more general implication of 
this observation of large fluctuations up
to 300 $\hmp$ for all standard analyses of existing 
catalogues is also clear.
If, on the other hand, the 
true underlying behaviour is fractal, the transition from
a highly fluctuating to a more stable behaviour can be understood
as the transition between large statistical fluctuations and
much smaller scale-invariant intrinsic fluctuations. In
Paper 1 it is shown how, using the dimension and 
normalization of the radial density from the analysis of 
the other surveys analyzed with the $\Gamma(r)$ statistic, 
one obtains a simple estimate for this scale in ESP which
agrees with the observation that it is $\sim$ 300 $\hmp$. 
The observed fluctuating regime is therefore consistent with 
the continuation of the fractal behaviour seen at smaller scales.
In fig.\ref{fig3} it is shown the distribution
of an homogeneous sample with the same geometry, number of 
points and luminosity selection effects of the ESP survey,
while in fig.\ref{fig4} it is shown the real ESP survey.
The fact that the survey is not homogeneous, or that
it does not become homogenous at some tens $\hmp$,  is evident by 
the simple visual inspection. 
\bef 
 \epsfxsize 7cm 
 \centerline{\epsfbox{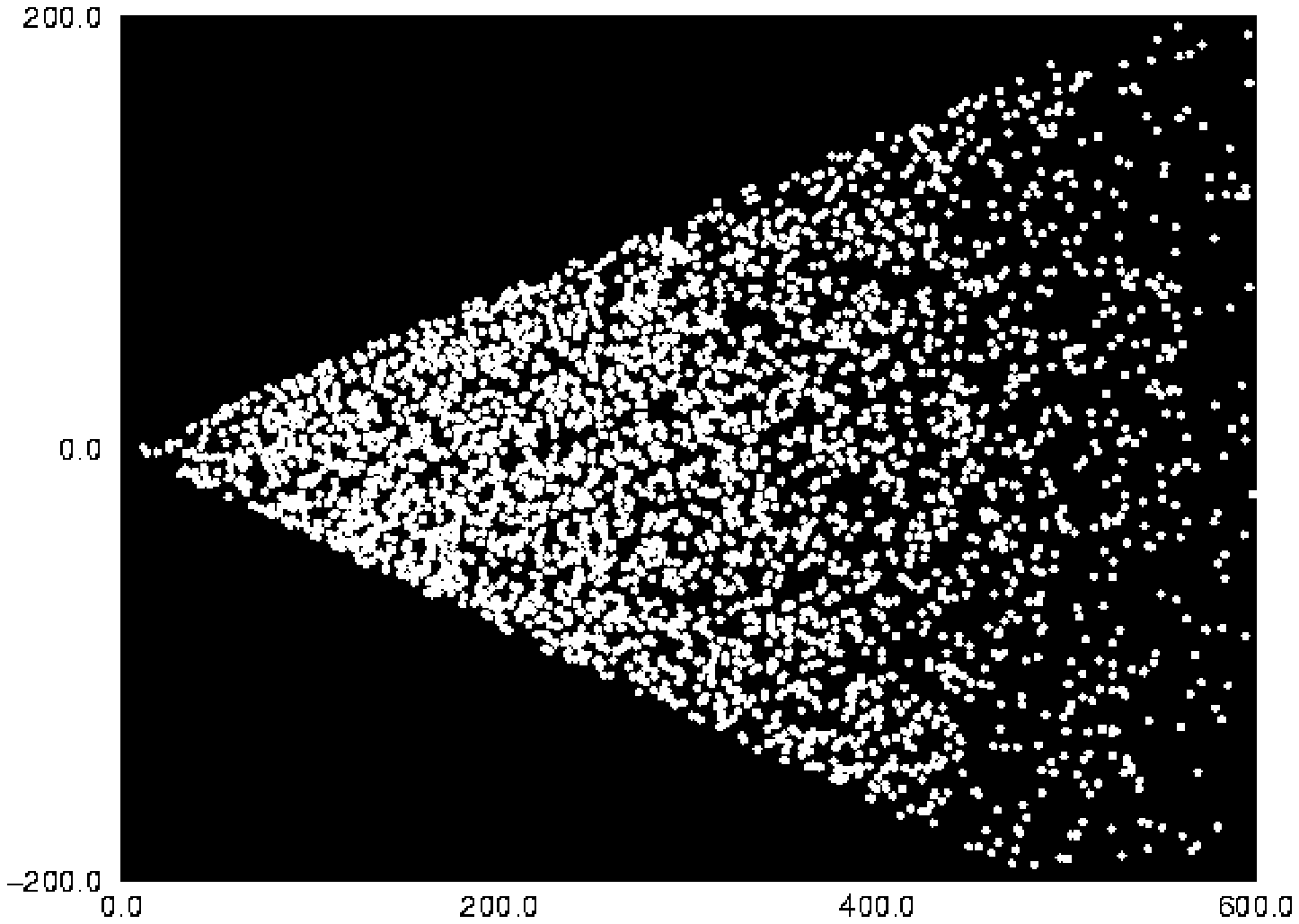}}  
 \caption{\label{fig3} 
 Distribution
of an homogeneous sample with the same geometry, number of 
points and luminosity selection effects of the ESP survey
in galactic coordinates} 
 \eef 
\bef 
 \epsfxsize 7cm 
 \centerline{\epsfbox{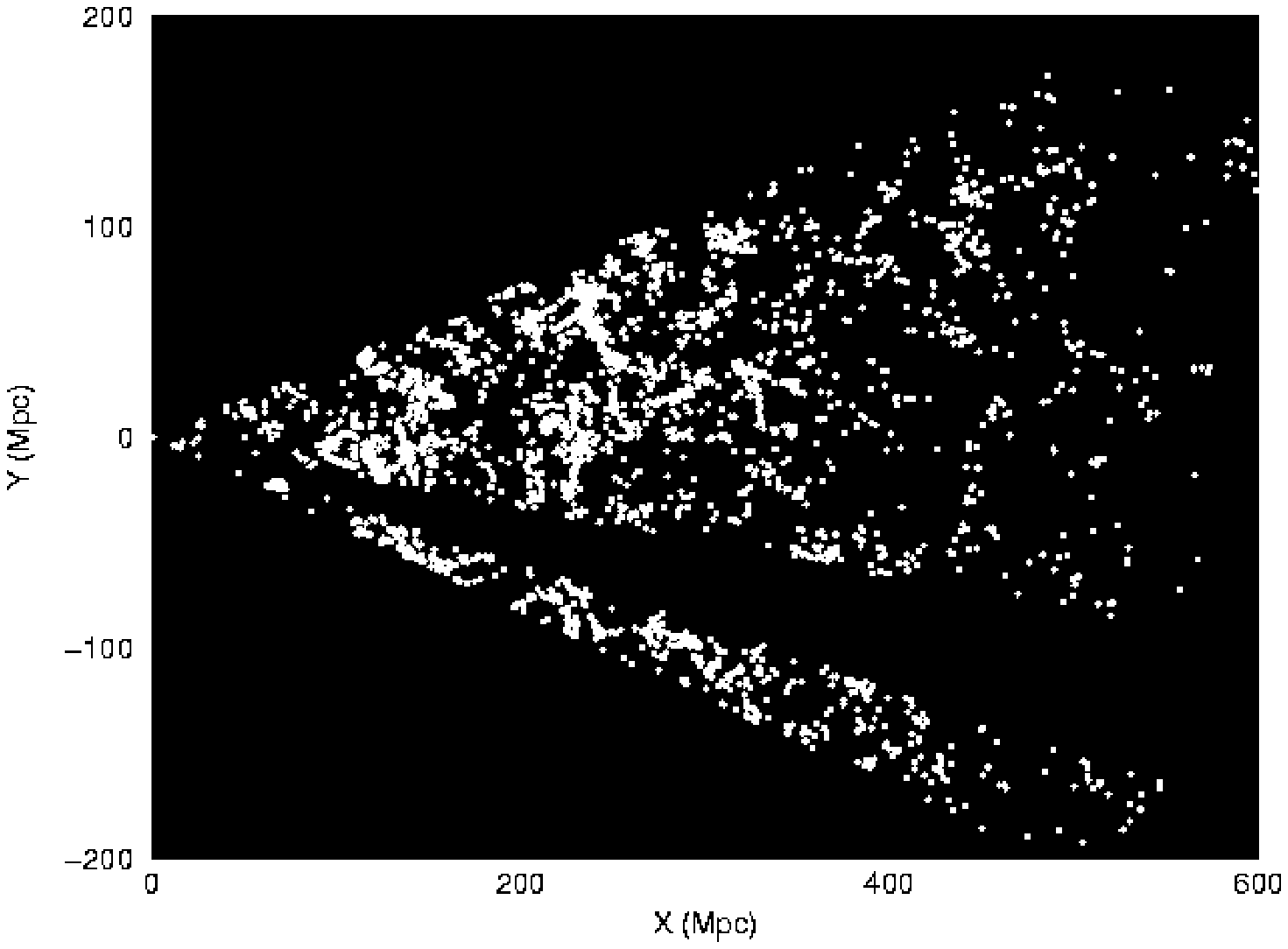}}  
 \caption{\label{fig4} The same as Fig.3 but for the real ESP survey 
 } 
 \eef

Now let us move onto the conclusions which can be drawn 
from the smoother regime at larger scales. In Paper 1
it was noted that from $ \sim 300 \hmp$ the number counts 
show very approximately a dimension $D \approx 2$,
and therefore provide weak evidence for the continuation
of the behaviour seen at smaller scales.
In S\&C the authors have looked at the precise effect which certain 
corrections can have on the data in the regime beyond $300 \hmp$
in much greater detail, and drawn the
strong conclusion that the $D \approx 2$ behaviour is ruled out
and $D=3$ behaviour clearly favoured. 
The difference of interpretation centers on the K corrections,
and the strength of any conclusion on the confidence with which
they can be made.  We will now see that some checking shows
that the K corrections as they have been applied to obtain this
result are in fact not just uncertain to an important extent,
but actually clearly inconsistent with the conclusion of
an underlying $D=3$ dimensionality. 

In order to construct VL limited catalogues we need to
determine the absolute magnitude $M$ of a galaxy at red-shift 
$z$ from its observed apparent magnitude $m$, and it is here that
the K correction enters:
\be
\label{abs-mag} 
M = m -5 log (d_L) -25 -K(z) \; .
\ee 
Here the luminosity distance $d_L=r(1+z)$
in FRW models where $r$ is the comoving distance,
and $d_L=(c/H_o)z$ in the euclidean case.
The K correction corrects for the fact that when a galaxy 
is red-shifted we observe it in a redder part of its spectrum 
where it may be brighter or fainter. It can in principle be determined
from observations of the spectral properties of galaxies,
and has been calculated for various galaxy types as
a function of red-shift.
When applying them to a red-shift
survey like ESP, we need to make various assumptions since
we lack information about various factors (galaxy types,
spectral information etc. as a function of red-shift
and magnitude).
 
It is not difficult to understand how K corrections
- whether themselves correct or incorrect - can change 
the number counts systematically. 
The number count in a 
volume limited (VL) sample corresponding to a magnitude limit 
of $M_{lim}$ can be written schematically as
\be
\label{number-count}
N( < R)= \int_0^R \rho(r) d^3r \int_{- \infty}^{M_{lim}}  \phi(M,r) dM 
\ee
where $\rho(r)$ is the real galaxy density, and $\phi(M,r)$ the 
appropriately normalized luminosity function (LF) in the radial
shell at $r$. If the latter integral is independent
of the spatial coordinate (i.e. if the fraction of galaxies
brighter than a given absolute magnitude is independent
of $r$), it is just an overall normalization and
the exponent of the number counts in the VL limited 
sample show the behaviour of the true number count, 
with  $N(<R) \sim R^D$ corresponding
to the average behaviour $\rho \sim r^{D-3}$ as we have assumed.
What we do when we apply K corrections or other red-shift 
dependent alterations to the relation (\ref{abs-mag}) is 
effectively change the LF $\phi(M,r)$. Using the incorrect relation 
will induce $r$ dependence in this function, which 
in turn will distort the relation between radial dependence of 
the number counts and the density. 

When one applies a K correction and observes a significant
change in the number counts, there are thus two possible
interpretations - that one has applied the physical
correction required to recover the underlying behaviour 
for the galaxy number density, 
{\it or} that one has distorted the LF to produce a 
radial dependence unrelated to the underlying density. 
How can one check which interpretation is correct? 
Consider the effect of applying {\it too large} a K correction of the
type applied by S\&C. To a good approximation
the net effect is a linear shift in the magnitude $M$ with 
red-shift, so that $M \rightarrow M - kz$.
The second integral in (\ref{number-count}) can in this
case be written 
\be
\label{shifted-integral}
\int_{-\infty}^{M_{lim}} \phi_p (M+kz) dM = 
\int_{-\infty}^{M_{lim}+kz} \phi_p (M) dM 
\ee
where $\phi_p$ denotes the physical, r-independent LF. 
This is a function whose shape is well known to
be fitted by a very flat power-law with an exponential
cut-off at the bright end. If the upper cut-off of the
integral is in the former range, one can see from 
(\ref{number-count}), taking $z \sim r$, that 
the number count picks up an additional
contribution going as $R^{D+1}$, so that 
we expect the slope to increase by one 
at some sufficiently large scale.
As we go to the brighter end of the luminosity function,
where it turns over, the fractional number of galaxies 
being added by the correction is even greater and we 
expect to see a growing effect on the slope with depth 
of sample. 
On 
the scales over which the ESP data are analysed, 
we should be able to clearly distinguish the case of
a spurious K correction from the case of a real
underlying $D=3$ behaviour, which should be relatively
stable as a function of the absolute magnitude limit of
the VL sample. In Figure 2 of their paper S\&C show the 
number counts for a few VL samples,
and conclude that there is evidence for a real sample slope
of $D=3$, the variation being ascribed to random errors.
In Fig. \ref{fig5} we show precisely the same figure with additional
VL samples at greater depth, which have been omitted for no apparent
reason by S\&C.
\bef 
\epsfxsize 6cm 
\centerline{\epsfbox{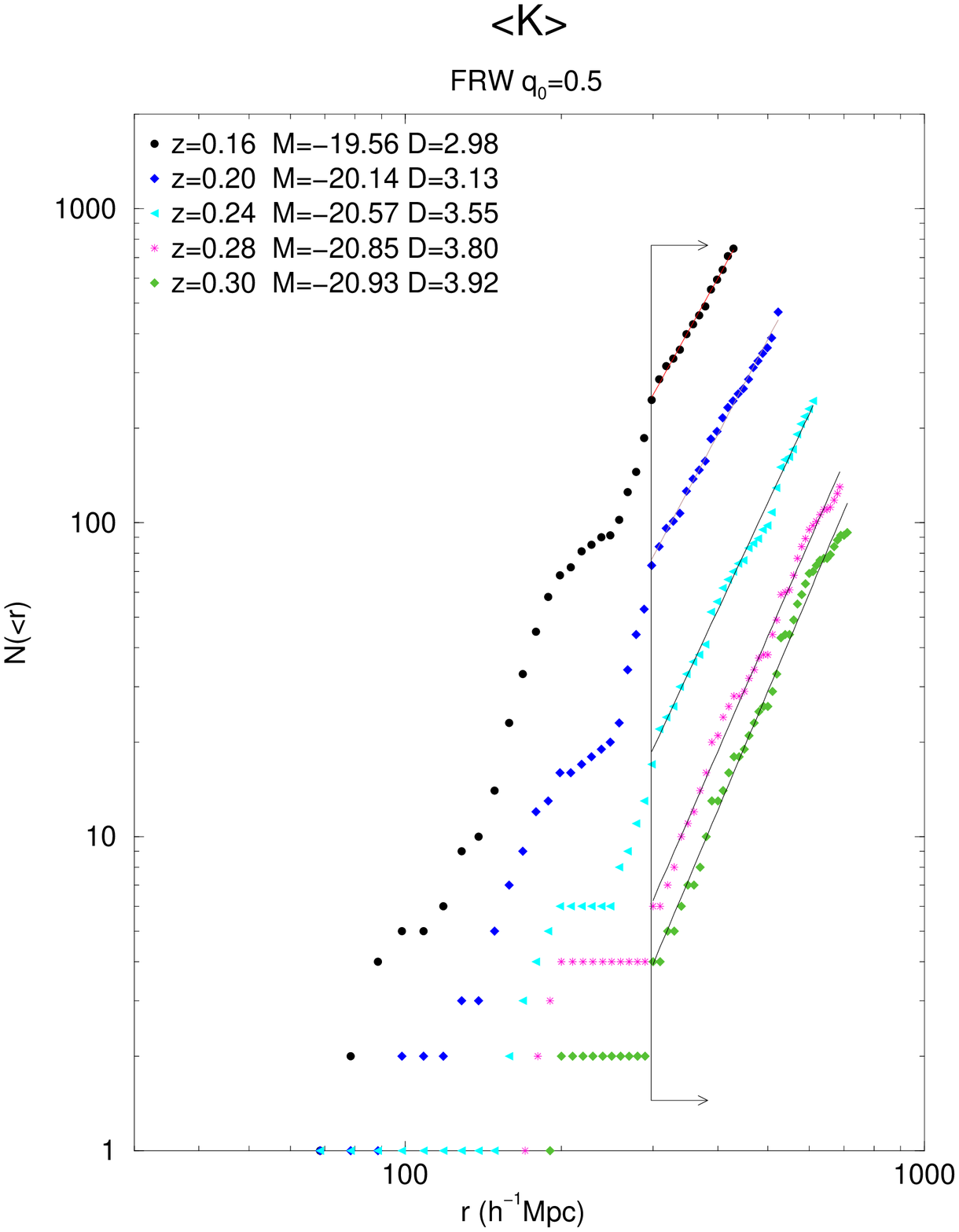}} 
\caption{ \label{fig5} 
The figure corresponds to Fig.2 of S\&C
($q_o=0.5$ with exactly the K correction given
in Zucca \etal  1997), but including deeper 
VL samples. The systematic growth of the exponent 
of the number counts as a function of the 
absolute magnitude limit of the sample is clearly
seen.} 
\eef
The conclusions one can draw from these two figures 
are clear: The $D\approx 3$ behaviour observed by
S\&C in their K corrected data is clearly
not stable as would be the case if it represented the
real underlying behaviour of the density. On the 
contrary the 
``corrected'' data are in fact clearly better interpreted
as indicating an underlying distribution with 
$D \approx 2$ which has been subjected to an 
unphysically large K correction in the relevant range
of red-shift $z \sim 0.1 -0.3$.

In contrast with the conclusion of S\&C
that the $D \approx 2$ result of Paper 1 is only 
tenable by  ``both unphysically ignoring the galaxy
K-correction and using euclidean rather than FRW cosmological
distances'', we conclude therefore that the alternative $D=3$ result 
is arrived at only by applying an unphysical K correction 
and taking a quite specific FRW cosmology.

It is perhaps interesting
to note that any approximately linear K correction will
not produce a stable dimension near $D=3$ for these data in these 
models. Clearly the physical K correction in this range 
of red-shift must be a very non-trivial function quite 
different from that used by S\&C if the ESP data arises from
an underlying homogeneous distribution. 
In the absence of a clearly consistent and well understood way of
applying such corrections, it makes little sense to draw conclusions 
which depend so strongly on them. By contrast the interpretation
of a continuation of fractal behaviour with $D\approx 2$ (and a
relatively unimportant role for K corrections) is a consistent 
interpretation, supported also by the behaviour seen in the 
fluctuating regime. With a smaller linear K correction with
$k=1$, for example, we find a stable slope in euclidean coordinates 
around $D=2.2$ (see fig.\ref{fig6}).
\bef 
\epsfxsize 6cm 
\epsfysize 8cm 
\centerline{\epsfbox{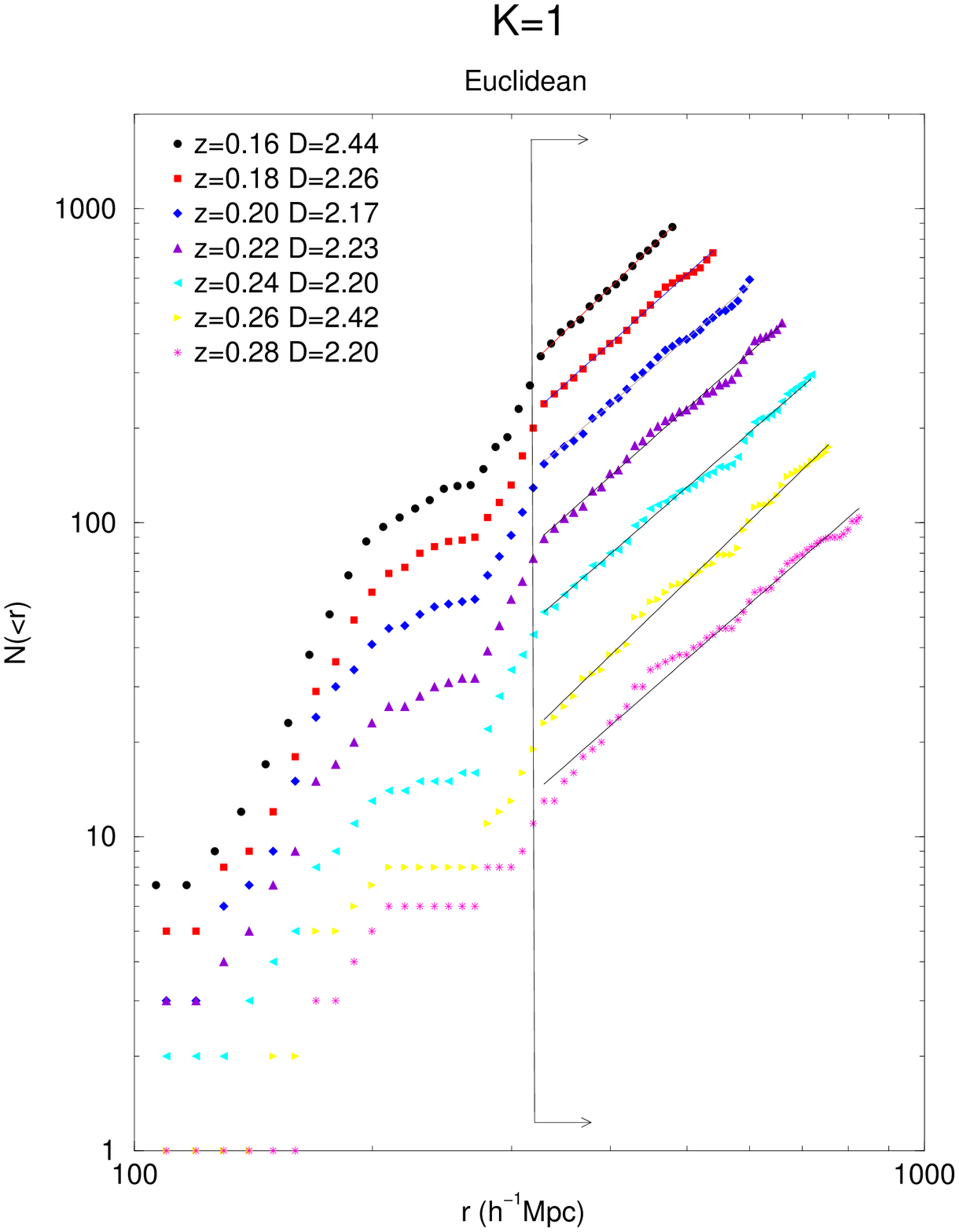}} 
\caption{ \label{fig6} 
In this case we have adopted a linear K-correction
$k=1$. We find a stable slope ($D\approx 2.2.$)
in euclidean coordinates.} 
\eef


\section{Uniformity of angular catalogs}

One of the most important elements in the discussion about 
galaxy correlations, is  
the analysis of angular distributions   \cite{dav97,pmsl97}.
Angular catalogs are
qualitatively inferior to  three dimensional  ones because they 
correspond to the angular projection and 
do not contain any information on the third coordinate. However,
the fact that they contain more galaxies than  
the 3-d
catalogs has led some authors to assign an excessive importance 
to these catalogs and  they are supposed to represent a clear 
evidence for homogeneity \cite{pee93,dav97,pee98,wu98}. 
Actually the interpretation of
angular catalogs is quite
delicate and ambiguous for a variety 
of reasons which are usually 
neglected \cite{cp92,slmp98,msl97}.
It is important to
stress that the existence of large scale structures
of galaxies has been found only in redshift surveys, while 
angular catalogs are relatively uniform. The 
reconstruction of 3-d properties of the galaxy distribution
from angular ones, is based on a series of assumptions 
that must be tested in real data. We show that the 
usual hypotheses used so far  contradict the behavior
found in the data analysis of angular correlations 
free of any a priori assumption.

Usually the analysis of angular correlations of galaxies 
is performed through the two point correlation function 
$\omega(\theta)$. This  allows one to determine 
a well defined characteristic scale in the angular distribution
(defined by $\omega(\theta_0) = 1$), and the correlation 
exponent at small angular separation is found to be 
$\gamma_a = 0.7$ \cite{pee80}. This value
of the power law exponent is claimed to be compatible
with the value $\gamma=1.7$ found in  3-d samples,
by the $\xi(r)$ analysis  \cite{pee80,pee93}.
In particular, the standard method used to analyze angular catalogs,
is based on the assumption that galaxies are correlated only at small
distances. In such a way  the effect of the large spatial inhomogeneities
is not considered at all. Under this  assumption, which is
not
supported by any observational evidence, it is possible to derive
Limber's equation  \cite{pee80}. In practice,
the angular analysis is performed by computing the two 
point correlation function
\be
\label{eeq1}
\omega(\theta) = \frac{\langle n(\theta_0)n(\theta_0+\theta) \rangle}
{\langle n \rangle} -1
\ee
where $\langle n \rangle$ is the average density in the survey.
This function is the analog of $\xi(r)$ for the 3-d analysis.
The results of such an analysis are quite similar to the 
three dimensional ones \cite{pee80,pee93}.
In particular, it has been obtained that, 
in the limit of small angles,
\be
\label{eeq2}
\omega(\theta) \sim \theta^{-\gamma+1}
\ee
with $\gamma \approx 1.7$ (i.e. $\gamma_a=\gamma-1=0.7$).
  
We now study the case of {\it a self-similar angular distribution}
so that, if such properties are present in real catalogs, we are 
able to recognize them correctly.
Of course, if the distribution is homogenous, we are able to
reproduce the same results obtained by the 
$\omega(\theta)$ analysis.
Hereafter we consider the case of small angles ($\theta \ltapprox 1$), that
is quite reasonable   for the catalogs investigated so far.
In this case the number of points within a cone of opening angle 
$\theta$ scales as 
\be
\label{eeq3}
N(\theta)= B_a \theta^{D_a} 
\ee
where $D_a$ is the fractal dimension corresponding to the 
angular projection and $B_a$ is related to the lower cut-off of the 
distribution. Eq.\ref{eeq3} holds from every occupied point, and 
in the case of an homogenous distribution we have $D_a=2$.
Following Coleman \& Pietronero \cite{cp92}
we define the conditional average density as
\be
\label{eeq4}
\Gamma(\theta)= \frac{1}{S(\theta)} \frac{dN (\theta)} {d\theta} = 
\frac{BD_a}{2\pi} \theta^{-\gamma_a}
\ee
where $S(\theta)$ is the differential solid angle element 
($S(\theta) \approx 2 \pi \theta$ for $\theta \ll 1$) 
and $\gamma_a=2-D_a$
is the angular correlation exponent (angular 
codimension). The last equality 
holds in the limit $\theta < 1$. From the very definition of 
$\Gamma(\theta)$ we conclude that
\be
\label{eeq5}
\omega(\theta)=\frac{\Gamma(\theta)}{\langle n \rangle} -1 \; .
\ee
A first important consequence of Eq.\ref{eeq5} is that 
if $\Gamma(\theta)$ has a power law behavior, and  $\omega(\theta)$ is
a power law minus one. This corresponds to a break in the log-log plot
for angular scales with $\omega(\theta) \ltapprox 1$. We show in 
Fig.\ref{figang1} the behaviour of such a quantity. 
\bef 
\epsfxsize 6cm 
\epsfysize 8cm 
\centerline{\epsfbox{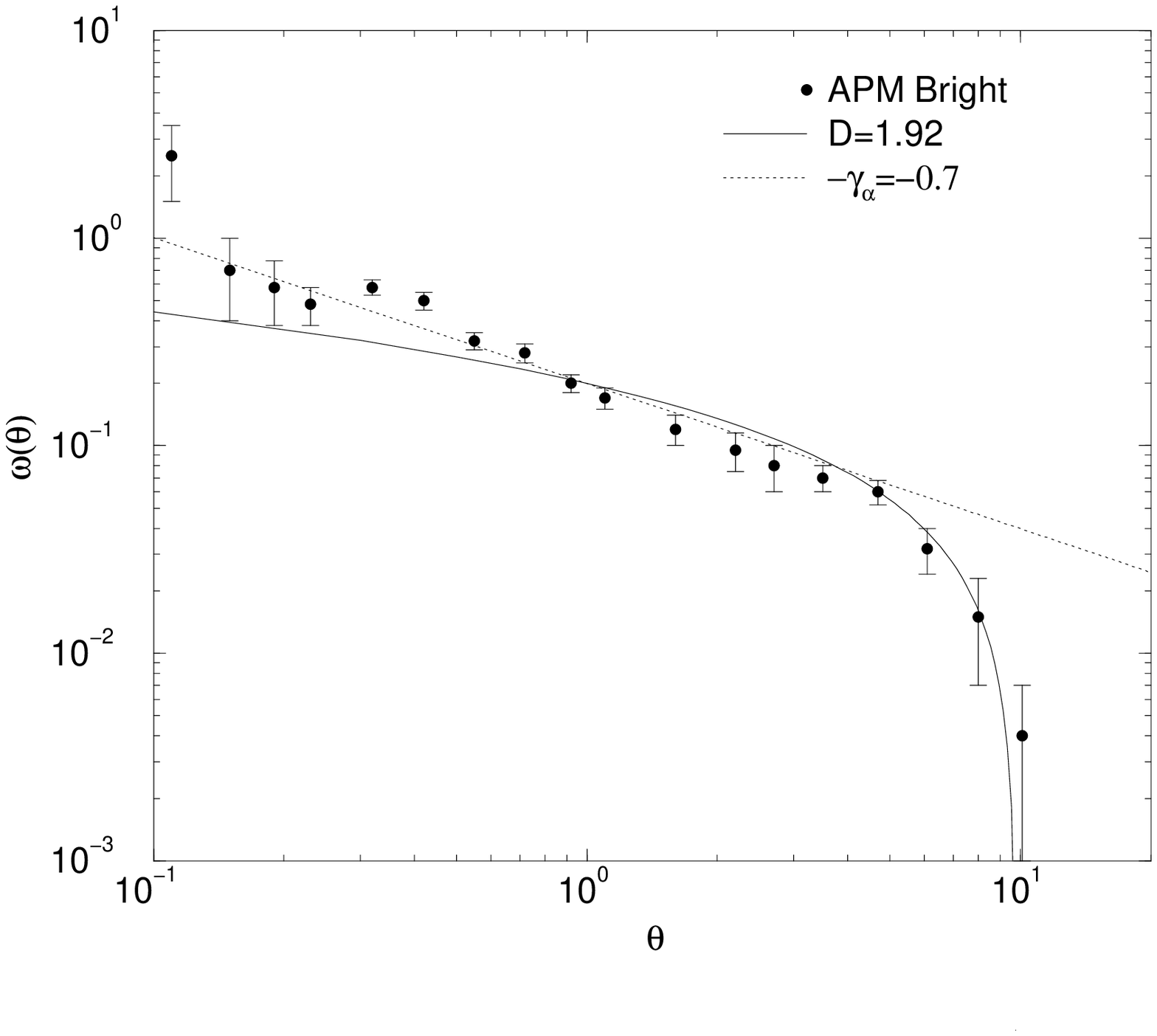}} 
\caption{In this figure we show the behaviour of
$\omega(\theta)$ (dotted line)
 is the case of a fractal structure with $D_a=1.92$
($\gamma_a=0.08)$.
It can be seen that the exponent obtained by fitting this function 
with a power law behavior (solid line) is higher than the real one 
($\gamma =-0.7$). Also the break in the power law behaviour is 
completely artificial. The amplitude has been matched to the
one of APM-BG with $m_{lim}=16.44$ (filled circles). \label{figang1} } 
\eef

The codimension found by fitting $\omega(\theta)$
with a power law function is higher than the real one. This is an important 
effect which
has never been considered before. 
In Fig.ref{figang1} we show also the $\omega(\theta)$ for the APM Bright Galaxies 
catalog (see below), 
that is fitted quite well by Eq.\ref{eeq5} with $D = 1.92$.
The second important point is that the break
of $\omega(\theta)$ in the log-log plot is clearly artificial and does not
correspond to any characteristic scale of the original distribution.
The basic 
problem is that in the case of a scale-invariant distribution the 
average density in eq.\ref{eq1} is not well defined, as it depends 
on the sample size \cite{cp92}.

Before we proceed, it is useful to recall the theorem for 
orthogonal projection of fractal sets.
Orthogonal
 projections preserve the 
sizes of objects. If an object of fractal dimension
$D$, embedded in a  space of dimension $d=3$, is projected on a plane
(of dimension $d'=2$) it is possible to show that 
the projection has dimension $D'$ with  \cite{man77,cp92} 
\be
\label{epro1}
D'=D \; \; \mbox{if} \; \; D<d'=2 \; ;
 \; D'=d' \; \; \mbox{if} \; \; D \ge d'=2 \; .
\ee
This explains, for example, why clouds which have fractal dimension
$D \approx 2.5$, give rise to a compact shadow of dimension $D'=2$.
The angular projection represents a more complex problem 
due to the mix of different 
length scales. Nevertheless 
the theorem given by Eq.\ref{epro1} can be extended 
to the case of angular projections in the limit of 
small angles ($\theta < 1$). Therefore according to 
Eq.(6) we have $D'=D_a$

We have analyzed the angular properties of the following catalogs:
CfA1, SSRS1, Perseus-Pisces,
Zwicky   and APM-Bright galaxies (APM-BG ).
The results are shown in 
Fig.\ref{angfig2}.
\bef 
\epsfxsize 6cm 
\epsfysize 8cm 
\centerline{\epsfbox{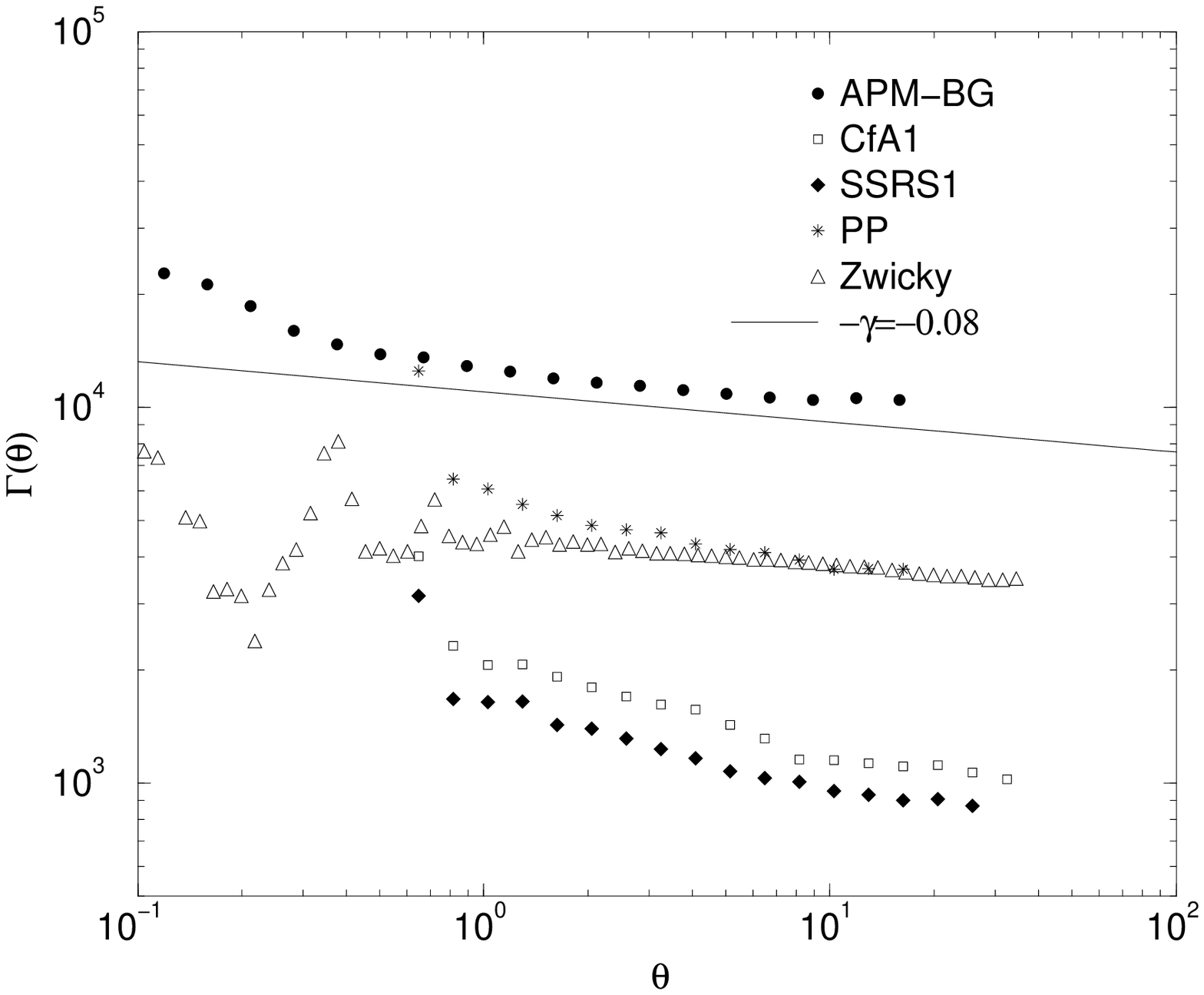}} 
\caption{ \label{angfig2} Angular correlation function $\Gamma(\theta)$
 for the magnitude limited 
samples CfA1, SSRS1, Perseus-Pisces (PP), Zwicky and 
APM Bright galaxies (APM-BG). 
The reference line has a slope $-\gamma_a=-0.08$ 
which corresponds to a fractal dimension $D=1.92$ in the 3-d space. 
The different amplitudes correspond to the different cut in apparent magnitude 
in the various catalogs} 
\eef
It turns out 
that all the catalogs show consistent correlation properties.
The angular 
fractal dimension is $D_a \approx 0.1 \pm 0.1$ 
(depending on the sample analyzed). No characteristic 
angular scale is present in any of the analyzed catalogs.

The angular distribution of galaxies turns out to exhibit marginal
scale invariance with angular fractal dimension
$D_a = 0.1 \pm 0.1$. Such a result, in view of the theorem
for orthogonal projections of fractal sets   is fully compatible
with the existence of a three dimensional fractal structure with dimension 
$D = 1.9 \pm 0.1$ which we have obtained in the 
analysis of the redshift samples
(Paper 1). This result alone is marginally 
compatible 
with an homogenous distribution in real space, because if $D >2$ than
we have $D_a = 2$. It results therefore that the angular analysis 
alone cannot be a {\it strong evidence} in favor of either 
a  homogeneous 
or  a fractal 
distribution in space with dimension 2. However, we stress again that 
the result $\gamma_a=0.7$ is just an artefact due to an inconsistent
 data analysis. 

It is useful to discuss briefly the angular 
 fluctuations expected in the 
case of fractal dimension $D$. It is possible to show\cite{pee93}
that the mean square fluctuations of the counts in two field 
of angular size $\theta$, with centers separated 
by angular distances $\Theta  \gg \theta$ is given by
\be
\label{flu}
\langle (N_1 - N_2)^2 \rangle \sim \langle N \rangle^2 
(\theta^{-\gamma_a}-\Theta^{-\gamma_a})
\ee
where $\langle N \rangle$ is the number of points over the whole sky
(it depends on the apparent magnitude limit of the sample).
If the value of the fractal dimension approaches two, then
$\gamma_a \rightarrow 0$ and the angular mean square 
fluctuations $\langle (N_1 - N_2)^2 \rangle \rightarrow 0$\cite{slm98}.
As we find $\gamma = 0.1 \pm 0.1$, this is compatible with a fractal
distribution in space with $D \approx 2$, and explains the 
uniform distribution of angular maps. A fractal distribution in space 
is characterized by having strong inhomogeneities on all scales, 
while the angular projection can be quite uniform and isotropic 
if $ D \gtapprox 2$, and  exactly this 
appears to be the case in  the galaxy catalogs (redshift and angular)
available up to now.

 Dogterom \& Pietronero \cite{dp91}   studied
 the surprising  and subtle  properties of the angular
 projection of a fractal distribution. They find that the angular
 projection produces an {\it artificial crossover towards homogenization
 with respect to the angular density}. 
 This crossover  is artificial
 (just due to the projection)  as  it
 does not correspond to any
 physical features of the three dimensional distribution. Moreover
 they showed that there is an explicit dependence of the angular
 two point correlation function $\omega(\theta)$ on $\theta_M$ the
 sample angle: this effect has never been taken into account in the
 discussion of real angular catalogs. These arguments show that it
 is very dangerous to make any definite conclusion just  from the
 knowledge of the angular distribution. By the way 
 this is the reason why one has to measure redshifts,  
 which, of course,  is not an easy task.

Let us now 
consider the much debated 
angular projection of an artificial fractal. Some authors
(e.g. Peebles \cite{pee93,pee98}) pointed out that a 
fractal with dimension $D$
significantly less than three {\it cannot}
 approximate  the observed 
isotropic angular distributions of deep
 samples. In particular Peebles 
\cite{pee98}
showed that a fractal with dimension 
$D \approx 2$ has  large-scale angular 
fluctuations which are not compatible with
 the observed angular maps. We   stress that there are various
problems, which are usually neglected  
in constructing an artificial distribution
with the properties of the real one:
\begin{itemize}

\item The first point is that in generating an artificial 
fractal structure a very important   role is
  played by {\it lacunarity}: 
 even if the fractal dimension
is fixed, one can have very different angular distributions
depending on the value of the lacunarity. 
Mandelbrot \cite{man77,man98} has insisted 
from a long time on this point.
In fact, if lacunarity is large,
and the sample is characterized by having voids 
of the order of the sample size, it is clear 
that the angular distribution is 
highly inhomogeneous. On the other
hand, if lacunarity is small 
(with respect to the sample size)
one can obtain more uniform angular 
projections .
A low value of the 
lacunarity should therefore be used for
the reproducing galaxy distribution, because the
real galaxy distribution has indeed a 
low value of the lacunarity:
in the available samples, the dimension 
voids is  smaller than the 
survey volume. A more detailed and realistic study
of this problem is now in progress.

\item
The second important point which should be considered is that the real angular
distributions are {\it magnitude limited ones}, i.e. contains all the galaxies
with apparent magnitudes brighter than a certain limit $m_{lim}$. 
This implies a mixing of
length scales due to the fact that galaxies have a very 
spread luminosity function, and 
their absolute luminosity can change of more than a factor ten. 
For example 
suppose that  $m_{lim}=14$, then one gets contributions from galaxies 
in the range of distances $\sim 1.5 \div 50 \hmp$.
However if, for example,   $m_{lim} = 17$ then the range of distances 
from which one has important contributions to the angular 
distribution, rapidly grows. 
This  implies that
 galaxies in angular catalogs correspond
to extremely different distances. Correspondingly 
the projection is a 
complex convolution of various 
luminosities and distances. 
This is another important 
reason why if one looks an angular 
map limited at galaxies
with apparent magnitude
 brighter than $14$ one sees large scale fluctuations. 
On the other hand a catalog 
(as the angular APM catalog) limited at $m_{lim} > 18$,  
is quite  smoother, because the mixing of length scales is large. 

\item An important point, recently pointed out by Durrer \etal \cite{durrer}, 
is the following: The angular lacunarity of a 3-d fractal set 
with $D \approx 2$ can be very small, 
even if there are large voids in the space distribution. In fact,
it can be shown that  the angular
lacunarity depends on whether the galaxies are projected with 
apparent or fixes size, being much more uniform in the latter case.

\end{itemize}

\section{Comparison of N-body simulations with data}
It is simple to see
that a fractal behavior of galaxy distribution
with dimension $D \approx 2$ up to, at least, $\sim  50 \hmp$
is not compatible with standard CDM models.
In fig.\ref{cdm}
\bef  
\epsfxsize 10cm 
\epsfysize 10cm 
\centerline{\epsfbox{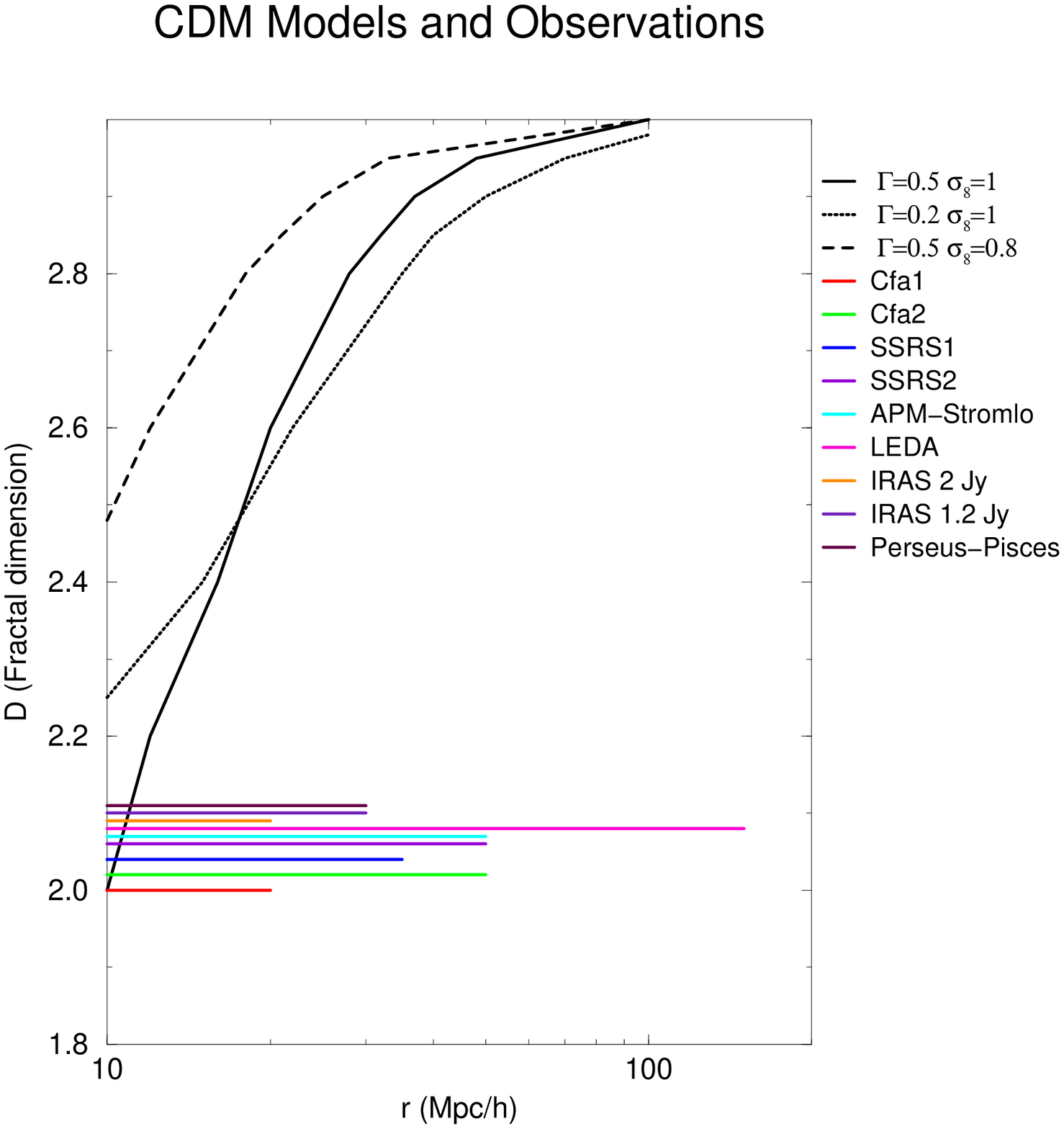}} 
\caption{\label{cdm} 
The fractal dimension versus distance in three
Cold Dark Matter models of power spectra 
with shape and normalized parameters as reported
in the labels (solid lines) (from Wu \etal, 1998).
We show also the different experimental determinations
of the fractal dimension we have obtained.
No agreement can be found at any scale. }
\eef
we show the behavior of the fractal dimension versus distance in three
Cold Dark Matter models of power spectra 
with shape and normalized parameters  (from Wu \etal, 1998).
We may see that a fractal dimension of 
$D \approx 2$ at $\sim 40 \div 50 \hmp$
is incompatible with all the models.
Probably by varying the parameters of
the simulation (or the mixture of Hot and Cold Dark Matter)
one may hope to obtain a better agreement.
Any new survey  has required a new adjustment
of the parameters  and
this alone shows the internal problems of the 
standard models of galaxy formation. 

We belive that the most important theoretical 
consequence of our results is that 
one may shift the attention
of the study from correlation amplitudes 
to  correlation exponents\cite{slmp98,durr98,sanchez1,sanchez2}.

\section{Conclusion}

We have tried to consider the main 
controversial points in the discussion 
about galaxy correlations.
The distribution of galaxies is one of the 
most robust experimental fact in modern
cosmology.
The new surveys like Sloan and 2dF
will explore a larger volume of the universe
than the one now available, and hence 
it will be possible to make and intensive study
of galaxy distributions up to $\langle z \rangle 0.1$.
We summarize our  opinion  
on this subject,   the corresponding technical discussion  
can be found in Paper 1. 
 
\begin{itemize} 
 
\item Until the seventies galaxy distributions were only known in terms of 
{\bf angular catalogues}. These catalogues are limited by the 
{\it apparent luminosity} of galaxies. Since the intrinsic 
luminosity can vary over a range of the order of one million, 
the points of the angular catalogues 
correspond to extremely different distances and they are a 
complex convolution of these - not just a projection up to some radius, which  
is the only case you discuss in your letter. These angular  
distributions appeared rather smooth and, as such, 
they  
justified the usual statistical assumptions 
of large scale homogeneity.
 
\item The redshift measurements permit to locate galaxies  
in 3-d space. They immediately showed a clumpy distribution with large clusters  
and large voids, in apparent contrast with the angular data.  
This finally led to 3-d catalogues from which one could make volume limited 
samples, which provide the best and most direct  
information  
for the correlation analysis. These 3-d data have been and are extensively 
analysed with the usual $\xi(r)$ method.  
The main result of this approach is that a characteristic length  
is derived $r_0 \approx 5 \hmp$ which should  
mark the tendency towards homogenization. This is in apparent 
agreement with the structureless angular data but it is puzzling  
with respect to the structures of the 3-d data. 
From this perspective it seems that the presence or absence  
of structures  
in the data is irrelevant for the determination of 
$r_0$.  
 
Puzzled by these results we have 
 decided to reconsider the question 
 of correlations from a broader and critical perspective. 
This allowed us to test homogeneity, 
instead of assuming it as in the $\xi(r)$ analysis, 
and to produce a totally unbiased description of the correlation properties  
using the methods of modern Statistical Physics.  
The main results are: 
 
\item The correlation properties of all the 3-d catalogues  
that we could collect are {\bf consistent with each other} and  
show fractal correlations up to the sample limits with dimension  
$D \approx 2$.  Such a situation allows us 
to establish the statistical validity and robustness
of the present redshift data. In this respect, we have also
performed several different tests to check whether 
the different catalogs are 
statistically stable and compatible with each other.
The result is that, a part several case (like the IRAS samples
which are very dilute), the optical catalogs are rather good.

\item The $\xi(r)$, with its a priori homogeneity  
assumption, appears therefore to be inconsistent  
for all these samples. The {\it "correlation length"} $r_0 \approx 5 \hmp$ 
is also an artefact of this analysis and $r_0$ is just  a fraction  
of the sample size. 
A consequence of   these results, is that  
the various catalogues 
appear to be in contrast with each other if the 
$\xi(r)$ method is used, but this apparent 
discrepancy is just due to the inappropriate statistics.

\item Also for the exponent (fractal dimension)  
the $\xi(r)$ analysis leads to a small value ($D \approx 1.2$ 
instead of the correct one $D \approx 2$)  because  
of its artificial drop in a log-log 
plot. 
It is interesting 
to note that the same problem  
for the determination of the correct exponent  
is also present in  the angular data where 
we get $\gamma \approx 0.1$ 
($D \approx 1.9$) instead of the usual
 value for $\xi(r)$ $\gamma \approx 0.7$ ($D \approx 1.3$). 
 
\end{itemize} 
 
  The clarification of the appropriate methodology for the analysis of the 3-d  
data is a very crucial point also in relation to the redshifts catalogues 
which will appear in the near future  (2dF and SLOAN). 
Only after this point is clarified it makes sense to consider the question  
of the quality of the data. In this respect we have made extensive tests 
on the effects of partial incompleteness   
and our conclusion is that the available samples are rather stable.
A strong support to our result comes also from the fact that  
the genuine correlation properties of the different catalogues 
are in good agreement with each other.
 
We have discussed at length various properties of  
the angular projections in previous papers. 
The problem of the angular data is that they are intrinsically 
ambiguous because  
one co-ordinate is missing and, in general, it is not possible  
to reconstruct (without assumptions)
 the properties of the real $3-d$ distribution 
from the angular catalogues. 
It is not a matter of number of points, it is a qualitative 
problem. In fact, if this would be possible,
 then there would be no need to measure 
the redshifts and to built $3-d$ catalogues.
 One can see instead that 
 all the large  
scale structures detected by   redshift measurements, could never be predicted  
by the angular data.  
 
So again, one cannot make any robust statement from the angular projection. 
In specific, for example, the projections depend on the value of the  
fractal dimension but also on a  variety of other properties like lacunarity,  
morphology and pixel size.  For example if the dimension is  
$D=2+ \epsilon$ it is very easy to have compact projections.  
If one has instead $D=2-\epsilon$  this is harder but, in any case, everything 
will depend on many other elements.  
 The simple example Peebles\cite{pee98}
 proposed is very far from a realistic analysis  
of these properties.  
 
In addition we find the  discussion on angular 
distribution rather confusing. For example,  
what has to do with our analysis the smooth 
projection of radio galaxies, that refers   
to sizes much larger than those we have ever considered ? 
The discussion has to address separate questions: 
\begin{itemize} 
 
\item[1] Are our results correct from a  
methodological point of view ? If yes, the previous results 
on the same data should be considered as incorrect  
on conceptual grounds. 
 
\item[2] The $3-d$ data available now may be incomplete or problematic. 
Then one has to wait for better data and analyse them anyhow with the methods 
we propose. The hope that the incompleteness of the data together with the  
conceptual mistake of the usual analysis may lead to correct results, should 
finally be abandoned.  Also the idea that some puzzling properties  
of the angular projections may magically save the usual scenario, is untenable. 
 
\item[3] How far the fractal properties extend ? 
This is an important question that has to be answered by new deep data. 
Our point was never to propose a model universe that  
is fractal all the way. We have instead proposed a new method of  
analysis that has disproved all the previously developed concepts  
($r_0=5 \hmp$ for galaxies, $r_0=25 \hmp$ for clusters, etc.) 
up to $100 \div 200 \hmp$, namely the region covered by  
extensive data (see also Teerikorpi \etal \cite{tee98}).
 If in the end the galaxy distribution will turn to be  
really homogenous at $\sim 2000 \hmp$ this will be detected with  
our methods and in no way will save the use and the  
results of the $\xi( r )$ approach.  
 
\item[4] The eventual fractal properties of galaxy distributions  
have nothing to do with the validity of General Relativity. 
The point is 
quite simple to define: we have experimental facts.
 These  are the galaxies with 
their correlations, the background radiation, 
the Hubble law, and various other facts.  
One should try to define these properties 
correctly and then make a theory that  
gives an understanding 
 for the properties of these various elements. 
If galaxies are more fractal than one would like, 
this maybe problematic for the models and theories
 of structure formation. Well, on should look for better theories. 
As for General Relativity, it is possible that is 
the end, the constant density assumption of Friedmann solutions should be changed into something more 
complex. This would not be the first time in Physics.
 In Condensed Matter only 
the Hydrogen Atom can be solved exactly. 
 
\end{itemize}

\section*{Acknowledgements}

We are in debt with Y. Baryshev, R. Durrer, 
M. Joyce, 
M. Montuori,
and P. Teerikorpi 
with whom various parts of these work have
been done. We warmly thank A. Amici, H. De Vega, 
H. Di Nella, J.P. Eckmann, 
A. Gabrielli, B.B. Mandelbrot, 
D. Pfenniger, 
N. Sanchez and F. Vernizzi 
for useful discussions and collaborations.
Finally we thank Prof. N. Sanchez and Prof. H. De Vega 
for their kind hospitality. 
This work has been partially supported by the 
EEC TMR Network  "Fractal structures and  
self-organization"  
\mbox{ERBFMRXCT980183} and by the Swiss NSF.


\begin{thebibliography}{99} 

\bibitem{pee93} Peebles, P.E.J., 1993 
Principles of Physical Cosmology, Princeton 
 Univ.Press, (1993) 

\bibitem{dac94}Da Costa \etal 1994 Astrophys. J. Lett 424, L1

\bibitem{dp83}Davis M. \& Peebles 1983, Astrophys. J.
267, 435

\bibitem{dav88}Davis M. \etal, 1988 Astrophys. J. Lett.  333 L9 

\bibitem{park94}Park C., Vogeley M. and Geller M.  
1994 Astrophys. J. 431, 569 

\bibitem{ben96}Benoist C., 1996 \etal  
Astrophys. J. 472, 452 

\bibitem{pie87}Pietronero L., 1987 Physica A, 144, 257

\bibitem{cp92}Coleman, P.H. and Pietronero, L.,1992 Phys.Rep. 231,311

\bibitem{slmp98} Sylos Labini F., Montuori M., 
 Pietronero L., 1998  Phys.Rep. 293, 66 (Paper 1)



\bibitem{man77} Mandelbrot, B.B., 
Fractals:Form, Chance and Dimension, W.H.Freedman, (1977) 

\bibitem{man98} Mandelbrot B., in the Proc. of the Erice Chalonge School
Eds. N. Sanchez and H. de Vega,  World Scientific  (1998) 
 
\bibitem{dav97}Davis, M.,  in the Proc of the  
Conference "Critical Dialogues in Cosmology" N. Turok Ed. (1997) 
World Scientific 

\bibitem{pmsl97} Pietronero, L., Montuori, M. and  
Sylos Labini, F., 1997 p.24-49 
in the Proc of the  
Conference "Critical Dialogues in Cosmology" N. Turok Ed.  
World Scientific 
 

\bibitem{gu97}Guzzo L., 1997 New Astronomy 2, 517 

\bibitem{slmp98b} Sylos Labini F., Montuori M., 
 Pietronero L., 1998a New astronomy submitted  
 (astro-ph/9801151)

\bibitem{coles98}Coles P., 1998 Nature 391, 120 

\bibitem{scara98}Scaramella \etal, 1998 Astronom.Astrophys. 
334,404 
 
 
 
\bibitem{cappi98}Cappi A., Benoist C., da Costa L.N. and Maurogordato S.
Astronom.Astrophys. 1998 335,779

\bibitem{joyce98}Joyce M., Sylos Labini F., Montuori M. and Pietronero L.
Astronom.Astrophys. 1998 Submitted (astro-ph/9805126)
  

\bibitem{wu98}Wu K.K.S., Lahav O. and Rees M., (1998)  Nature submitted
(astro-ph/9804062)

\bibitem{pee98}Peebles, P.E.J., 1998 In 
the Proceed. of the Conference "les Rencontres de Physique de la
Vallee d Aosta" (1998) ed. M. Greco (astro-ph/9806201)




\bibitem{sanchez1}de Vega H., Sanchez N. and Combes F., (1996)  Nature 383, 53


\bibitem{sanchez2}de Vega H., Sanchez N. and Combes F. (1998) , Ap. J In Press

 

\bibitem{pee80} Peebles, P.E.J., 1980 "The Large Scale Structure of 
The Universe" 
(Princeton Univ.Press.); 


\bibitem{lov96}Loveday J. \etal , 1995 Astrophys. J. 442, 457 

\bibitem{lcrs}Schectman S. {\em et al.}, 1996
Astrophys. J.,   470,172
 

\bibitem{slm98}Sylos Labini F., Montuori M.,  1998 
Astron. Astrophys    331, 809   


\bibitem{msla97} M. Montuori, F. Sylos Labini  and A. Amici 1997 
Physica A,   246 , 1-17 



\bibitem{mslgpa97} Montuori M., Sylos Labini F. 
Gabrielli A., Amici A. and Pietronero L. 
1997 Europhys Lett. 31, 103 

\bibitem{slp96}Sylos Labini F., Pietronero L., 1996 Astrophys. J. 469,28

\bibitem{durr98}Durrer R. and Sylos Labini F. (1998) Preprint
(astro-ph/9804171)

\bibitem{bar98}  Y. Baryshev,   F. Sylos Labini, M. Montuori ,  
L. Pietronero, and P. Teerikorpi Fractals (1998) In the Press.


\bibitem{guzzo91}
Guzzo L. \etal , 1991 Astrophys. J. Lett., 382 L5 
 
\bibitem{zucca97}Zucca E., \etal., 1997 Astron.Astrophys. 326,477  

\bibitem{dp91} Dogterom, M., Pietronero,  L., 1991 Phys. A 171,  239

\bibitem{durrer}Durrer R., Eckmann J.-P.,  Sylos Labini F., Montuori M.,  
Pietronero L. 1997 Europhys. Lett. 40, 491 

\bibitem{msl97} Montuori M.  and  Sylos Labini F. 
1997 Astrophys. J. Lett. 487,  L21 
 
\bibitem{tee98} 
 Teerikorpi, P., \etal Astron.Astrophys.  In The Press (1998); astro-ph/9801197 

\end{thebibliography}
\end{document}